# Simulation of Open-loop Plasma Vertical Movement Response in the Damavand Tokamak Using Closed-loop Subspace System Identification

**N. Darestani Farahani** [a] **and F. Abbasi Davani** [a,*]

[a] *Department of Radiation Application, Shahid Beheshti University, Tehran, Iran*
*E-mail*: fabbasi@sbu.ac.ir.

ABSTRACT: A formulation of a multi-input single-output closed-loop subspace method for system identification is employed for the purpose of obtaining control-relevant model of the vacuum-plasma response in the Damavand tokamak. Such a model is particularly well suited for robust controller design. The accuracy of the estimate of the plant dynamics is estimated by different experiments. The method has been described in this paper is a worst-case identification technique, that aims to minimize the error between the identified model and the true plant. The fitness of the identified model around defined operating point is more than 90% and compared with physical-based model it has better root mean square measure of the goodness of fit.

## Contents

## 1. Introduction

At present, tokamaks (toroidal vessels with magnetic coils) [1] are the most promising devices for magnetic plasma confinement to achieve conditions of thermonuclear fusion. In tokamak devices, one method for increasing the viability of the tokamak as a fusion energy device is to operate with more strongly non-circular plasma cross-sections. In this way the current carrying capability can be significantly enhanced, leading to better energy confinement. However, increasing plasma elongation increases the risk of vertical displacement events (VDE) [2]. It is clear that the number of VDEs must be limited as much as possible, as a part of tokamak design criteria, which allow for steady-state operation [3].

The requirements of stable and advanced modes of tokamaks operation motivate researchers to improve the modeling of the plasma response as well as the design of feedback controllers. As an alternative to first-principles modeling, data-driven modeling methods [4] have been successfully used to develop plasma equilibrium. In the JET tokamak, a two-time-scale linear system has been used to describe the magnetic and kinetic profiles around certain quasi-steady-state trajectories [5]. H-infinity system identification technique was used to create a linear model of the response of the TCV tokamak plasma to voltages applied to the poloidal field coils[6].

In discharges at the JT-60U tokamak, the momentum transport equation of the toroidal rotation profile has been estimated from transient data obtained by modulating the momentum source [7]. System identification experiments have been carried out on the DIII-D tokamak [8, 9]. A nonlinear model has been identified for plasma vertical position in the Damavand tokamak, based on the multilayer perceptron (MLP) neural network (NN) structure [10] and a fractional order model has also been identified using output-error technique in time domain[11, 12].

Damavand tokamak is a small size tokamak and like other small size tokamaks in the world is a suitable device to perform research on fusion related subjects such as plasma diagnostics [13-16], control feedback systems and power electronics [10-12]. In the Damavand tokamak, hydrogen plasma is sustained for 21 ms with a plasma current peak of about 35 kA [10]. The plasma column undergoes hydrodynamic and magnetic pressures which ultimately reach magnetohydrodynamic (MHD) equilibrium in a stationary state [17] .

Damavand vertical stabilization is done with passive coils and active feed-back external coils (Cz). So, the passive coils effects on vertical displacement and the requirements of the Cz power supply (PS) are important for Damavand performance and require evaluation. An analog proportional + derivative (P+D) system is currently used for the vertical position control system on the Damavand tokamak. Design and implementation of different controllers based on black box modeling that is obtained from system identification techniques to control VDE in the Damavand tokamak, have been done and the controllers are implemented on a digital signal processor and tested on Damavand. The controller synthesis procedure and the experimental results are presented in [10-12].

This work is motivated by the need to create linear models which are suitable for controller design and which may be used to validate different models derived from physics principles. In this paper we describe how a subspace system identification technique was used to create a linear model of the response of the Damavand tokamak plasma to voltages applied to the poloidal field coils. In this study we were particularly interested in the dynamics from the PF coils to a parameter which describe the plasma vertical position. The plasma-coil system is Multi Input Single Output (MISO). Tokamak models, either identified or physical based, will play a key role in the design of control systems for future machines.

In Section 2 we detail the identification algorithm for a MISO continuous-time system. The identification procedure uses data collected from a series of identification experiments. We use a Pseudo Random Binary Signal (PRBS) which is applied to the inputs of the closed-loop system with signal frequencies selected using a bilinear formula. The vertical position of plasma is measured with respect to the center of the chamber by saddle loop sensor. In Section 3 we outline a case study performed on the Damavand tokamak. We detail the experiments which, due to the plant instability, were performed in closed loop and show how the method outlined in Section 2 is used to create identified open-loop model of the machine. This identified model is compared with experimental result and with a physical-based model result. Finally, in Section 4, conclusions are presented.

## 2. Identification Method

Because of the plasma vertical position instability in the Damavand tokamak, a suitable method for identification plasma response from experimental data would be the direct closed loop method. In this method, process should be controlled in a closed loop structure while the input and output data of the process are employed for identification of plasma response. Due to high speed processes the model is identified offline and is based on experimental data of the machine.
the problem of system identification is specified by three elements [18]: A data set obtained by input-output measurements , a model structure and a rule to evaluate candidate models, based on the data (model validation).

## 2.1 Experiment Design

This section includes three parts: First, magnetic field coils arrangement and coils structure of the Damavand tokamak is presented. Second, plasma vertical position control system in the Damavand tokamak is presented. Third, design and execution of experiments for identification is explained.

### 2.1.1 Damavand Tokamak

All the experiments have been done on the Damavand tokamak. As it was mentioned in the introduction, the Damavand tokamak is one of the small size tokamaks which is able to produce elongated plasma with different measures of elongation and triangularity. Toroidal magnetic field of the Damavand tokamak is up to $1.2\ T$ and the plasma duration is around $21\ ms$. Figure 1 shows the Damavand tokamak coils structure, their type and application in duration of plasma [19]. Capacitor banks are used for supplying energy for magnetic fields in the Damavand tokamak. The supplied energy for magnetic fields production are: toroidal magnetic field bank: 850kJ, central solenoid: 250kJ, equilibrium magnetic field: 35kJ, and feedback system: 5kJ. In order to measure the current of the coils, calibrated Rogowski coils which are routinely recalibrated are used. The coil current direction can be changed in different coils according to their functions. Data of diagnostics, pickup coils, saddle loops and coil currents are acquired by the data accusation system with sampling rate 2MS/s to 20 MS/s. The coil currents in the Damavand tokamak are measured with error up to 4%.

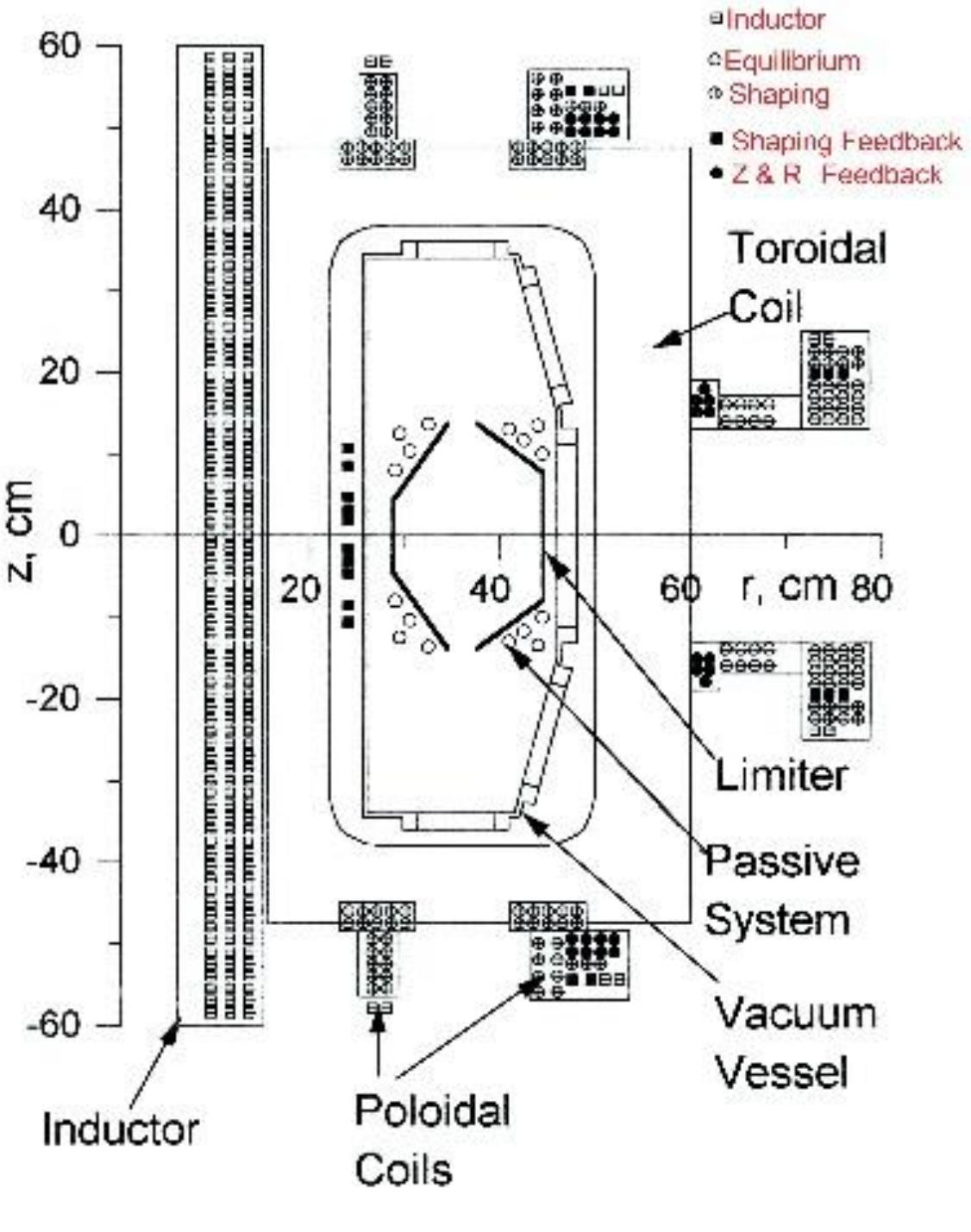

FIG. 1. Cross Section and Coils Structure of the Damavand tokamak

Figure 2 shows the plasma current and poloidal coil currents for production of plasma in an elongated cross section.

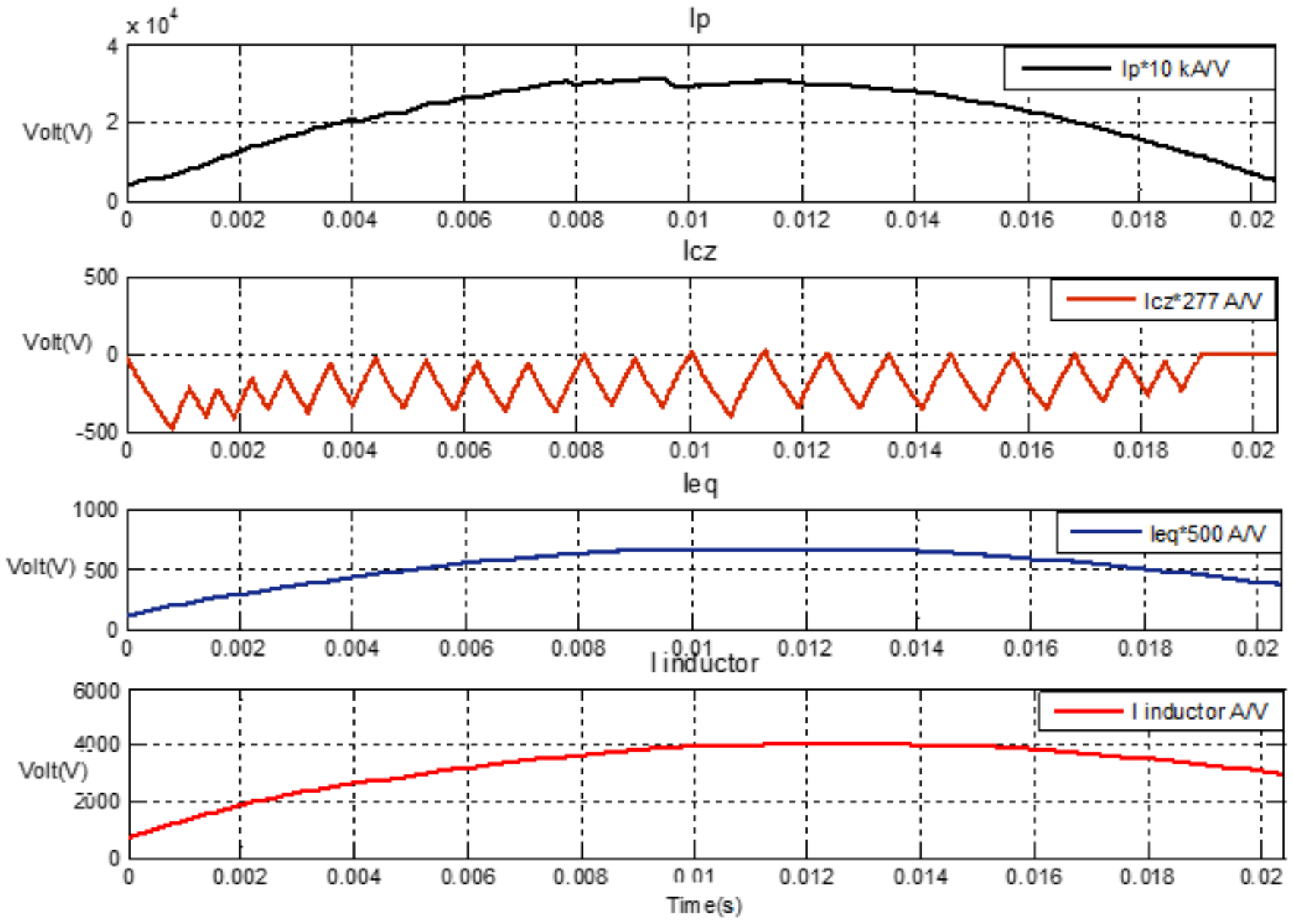

FIG.2. (a) Plasma current, (b) Vertical position control coil current, (c) Equilibrium coil current and (d) Inductor coil current for production of elongated plasma with zp=0. (2015/27/05 Shot#11)

### 2.1.2 Plasma Vertical Position Control System in the Damavand Tokamak

In the Damavand tokamak, the vertical position of plasma is measured with respect to the center of the chamber. The block diagram of closed loop system of plasma vertical position for the Damavand tokamak is shown in Figure 3. Here, $\dot{\psi}_z$ is the output of the sensor (saddle loop) and is fed into a circuit with transfer function of H(s) which in turn reconstructs the vertical position [10].

$$\mathbf{H(s) = k\frac{1010}{s + 1.377}} \quad (1)$$

$$\mathbf{G_c(s) = k_p + \frac{k_d S}{s + 10^4}} \quad (2)$$

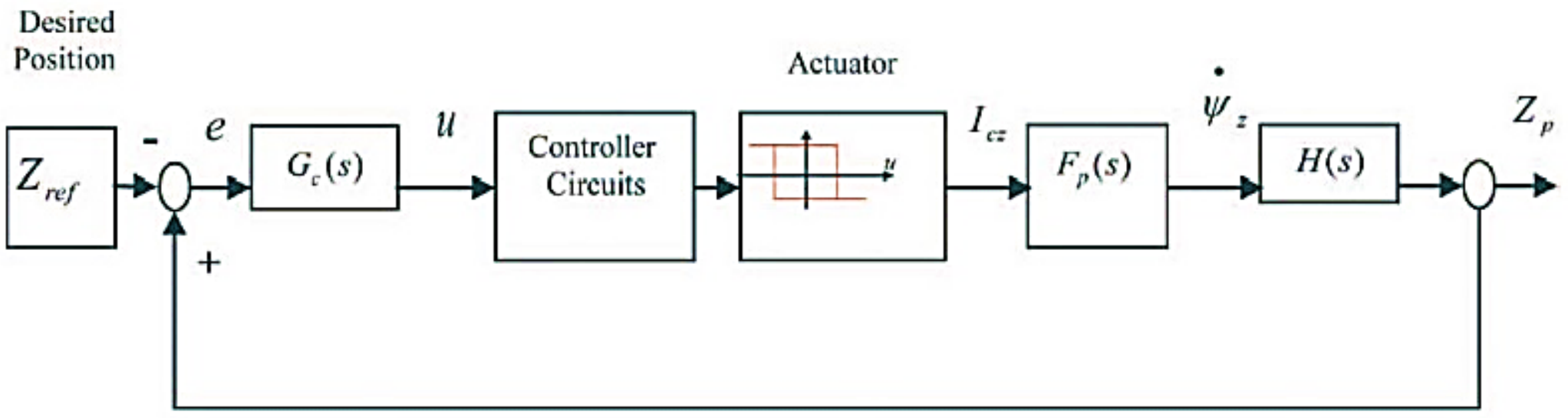

FIG. 3. Block diagram of closed loop system to control plasma vertical position in the Damavand tokamak.

In Figure 3, $Gc(s)$ is the controller with transfer function of relation(2) [10]. Also, $Fp(s)$ represents dynamic of the plasma vertical position. Using the control signal $(u)$ and the rectifier power circuit (actuator), the controller current $(Icz)$ flows into the external poloidal coils. The produced signal of block $Zref$ is used in closed loop system to set desired vertical position of the plasma column inside the vacuum chamber, ranging from $-12\ cm$ to $+12\ cm$[10]. According to Figure 3, $Fp(s)$ represents dynamics of the vertical displacement of the plasma.

### 2.1.3 Design and Execution of Experiments

The first step in system identification is designing a set of suitable experimental tests. The important subject in such designing is selection of input signal ($Zref$) which could excite all modes of the plant. The selected input signal is a Pseudo Random Binary Signal (PRBS) having a pulse length in the range of 1 $ms$–20 $ms$ with maximum amplitude of ±6V. Control structure of the system (shown in Fig. 3) was applied in the Damavand tokamak. In this structure, main variable parameters of tokamak were kept constant in all tests. Duration of each discharge is no longer than 20 $ms$. So, during this short time, identification of the plant (by) using data of just one shot is impossible. It is required to use different parts of identification signal gathered from various shots [10]. Figure 4 illustrates data gathered from different shots for nonlinear system identification (neural network) that used in previous work [20]. The aim of this work is to identify linear model for system. And this is possible just for fixed operating points. So different shots with fixed z-position gathered and identification process has been done for a single shot and validated for others. More explanation about the method and results are given in next sections.

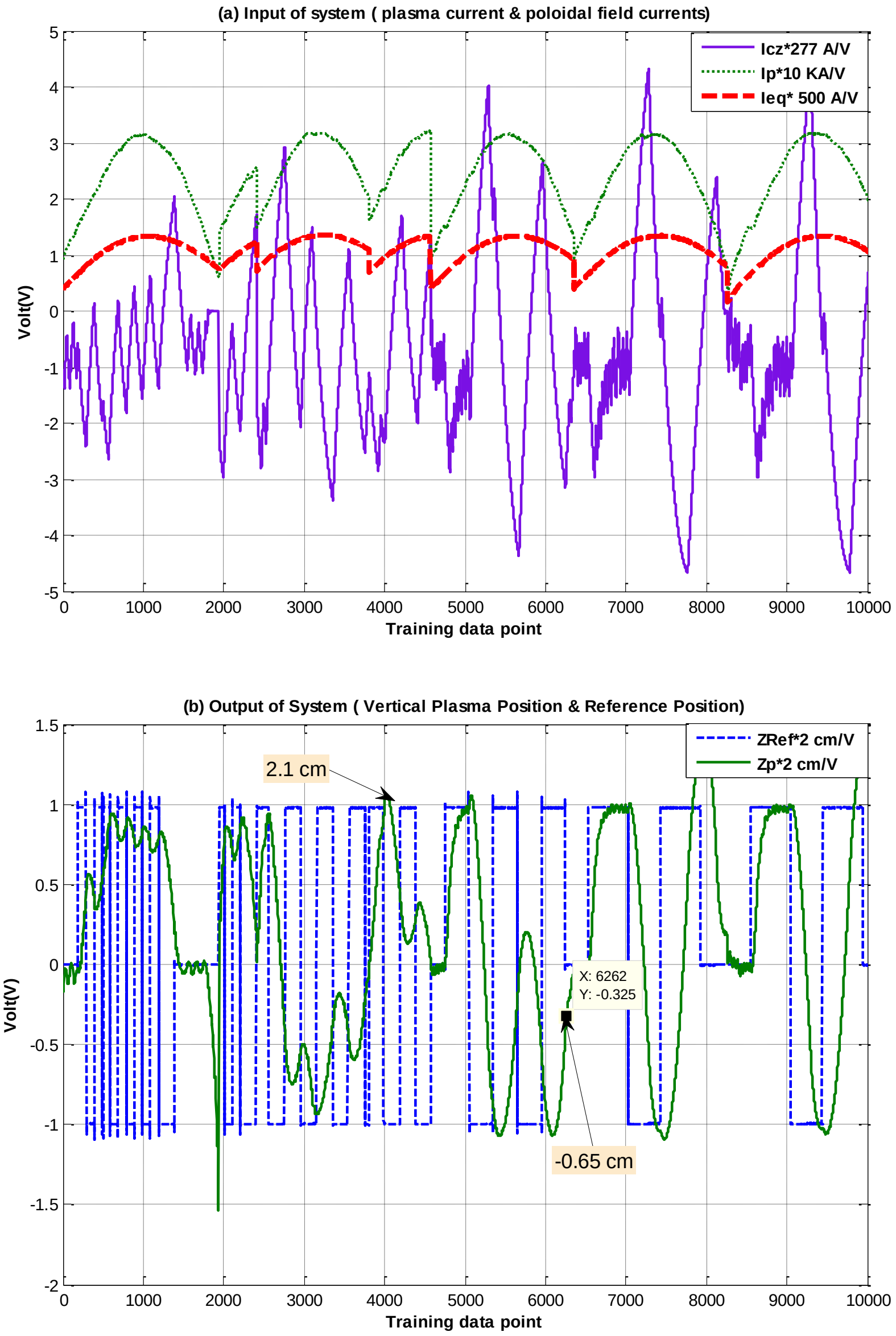

FIG. 4. A sample of shots for nonlinear model identification. (a) Input of system. (b) Output of system

Because of high frequency noise and signal type, pre-processing is required to eliminate noise, to correct bias and to select proper time range of input and output signals[10]. After data preparation in database, different models can be identified and evaluated. In the Sec.3, we will describe linear simulator model of the identified Fp(s).

## 2.2 Discrete-Time Linear Time-Invariant System Model

The determination of control laws for any system requires the knowledge of its dynamic model. The modelling involves then, for the system to be controlled, to obtain a mathematical model that well describes its dynamic behaviour. The system identification is the main powerful technique for building accurate mathematical models of complex systems from noisy data. Indeed, many identification approaches are studied in literature such as those called predictor error methods [21, 22] and those based on the approximate least absolute deviation criteria [23, 24]. The iterative nature of optimization algorithms used in these methods can lead to some problems like numerical instability or lack of convergence. The subspace identification techniques [25-28] constitute a good alternative to these classical methods, and especially for multiple-input and multiple-output (MIMO) linear systems. During the last decades, subspace techniques have been well used for systems identification. The main advantage in these new identification methods is that some nonlinear optimization algorithm is required [27]. Based on geometric and mathematical procedures, they use only robust tools of linear algebra, such as the QR decomposition (Q: orthogonal matrix; R: upper triangular matrice), the singular value decomposition (SVD) singular value decomposition, the matrices projection and the angles of subspaces [27]. The purpose of the subspace identification method is to estimate a linear time invariant state space model, directly from input-output measures. Recently, many researches work on subspace identification algorithms have been developed to estimate model parameters of industrial processes [29]. Among these algorithms, we can cite the numerical subspace state space system identification (N4SID) algorithm.

In this section, we are interested in the presentation of the mathematical and geometric tools used in subspace identification (N4SID) [30].

Given q measurements of the input $u_k \in R^m$ and the output $y_k \in R^l$ generated by an unknown deterministic system of order $n$ and described by the following discrete equations:

$$\begin{aligned} x_{k+1} &= Ax_k + Bu_k \\ y_k &= Cx_k + Du_k \end{aligned} \qquad (3)$$

Where $x_k \in R^n$ denote the state vector at time $k$. One should determine the realization $(A, B, C, D)$ with $A \in R^{n\times n}$, $B \in R^{n\times m}$, $C \in R^{l\times n}$ and $D \in R^{l\times m}$ that the unknown matrices to be identified.

Let $U$ be a vector such as $U = [u_0 \quad u_1 \quad \cdots \quad u_{q-1}]^T$. With $q = 2i + j - 2$ and $i < j$.

The Hankel matrices of past and future inputs are defined respectively as:

$$U_p = \begin{bmatrix} u_0 & u_1 & \cdots & u_{j-1} \\ u_1 & u_2 & \cdots & u_j \\ \vdots & \vdots & \cdots & \vdots \\ u_{i-1} & u_i & \cdots & u_{i+j-2} \end{bmatrix} \qquad (4)$$

$$U_f = \begin{bmatrix} u_i & u_{i+1} & \cdots & u_{i+j-1} \\ u_{i+1} & u_{i+2} & \cdots & u_{i+j} \\ \vdots & \vdots & \cdots & \vdots \\ u_{2i-1} & u_{2i} & \cdots & u_{2i+j-2} \end{bmatrix} \qquad (5)$$

In the same way, the Hankel matrices of past and future outputs are defined respectively as:

$$Y_p = \begin{bmatrix} y_0 & y_1 & \cdots & y_{j-1} \\ y_1 & y_2 & \cdots & y_j \\ \vdots & \vdots & \cdots & \vdots \\ y_{i-1} & y_i & \cdots & y_{i+j-2} \end{bmatrix} \qquad (6)$$

$$
Y_f = \begin{bmatrix} y_i & y_{i+1} & \cdots & y_{i+j-1} \\ y_{i+1} & y_{i+2} & \cdots & y_{i+j} \\ \vdots & \vdots & \cdots & \vdots \\ y_{2i-1} & y_{2i} & \cdots & y_{2i+j-2} \end{bmatrix} \tag{7}
$$

The instrumental variable matrix (or Hankel matrix of past data) is given by

$$
Z_p = (U_p \;\; Y_p)^T \tag{8}
$$

The matrix related to the sequence defining past $X_p \in R^{n\times j}$ and future $X_f \in R^{n\times j}$ :

$$
\begin{aligned} X_p &= [x_0 \quad x_1 \quad \cdots \quad x_{j-1}] \\ X_f &= [x_i \quad x_{i+1} \quad \cdots \quad x_{i+j-1}] \end{aligned} \tag{9}
$$

The extended observability matrix $\Gamma_i \in R^{li\times n}$:

$$
\Gamma_i = \begin{bmatrix} C \\ CA \\ CA^2 \\ \vdots \\ CA^{i-1} \end{bmatrix} \tag{10}
$$

$H_i$ is the block Toeplitz matrix.

$$
H_i = \begin{bmatrix} D & 0 & \cdots & 0 \\ CB & D & \cdots & 0 \\ CAB & CB & \cdots & 0 \\ CA^2B & CAB & \cdots & 0 \\ \vdots & \vdots & \cdots & \vdots \\ CA^{i-2}B & CA^{i-3}B & \cdots & D \end{bmatrix} \tag{11}
$$

The inputs excitation is persistent of order $2i\big(rank\ U_{0\,|2i-1}U_{0\,|2i-1}\big) = 2mi)$. Intersection between the row space of matrix $U_f$ and the one of matrix $X_p$.

One of the mathematical tools of linear algebra used in subspace method is oblique projection [27].With $O_i$ as the oblique projection:

$$
O_i = {}^{Y_f}\!\big/_{U_f}\, Z_p = \begin{bmatrix} Y_f \\ \hline U_f{}^{\perp} \end{bmatrix} \begin{bmatrix} Z_p \\ \hline U_f{}^{\perp} \end{bmatrix}^{\dagger} Z_p = Y_f \Pi_{U_f{}^{\perp}} + (C\Pi_{U_f{}^{\perp}})^{\dagger} Z_p \tag{12}
$$

Where the dagger sign (†) denotes the pseudo inverse Moore-Penros of matrix, **Π** is the operator of orthogonal projection which projects the row space of a matrix onto the row space of the matrix ($\Pi_{U_f} = U_f{}^T (U_f U_f{}^T)^{\dagger} U_f$).

We multiply $O_i$ at left and at right, respectively, by the matrices $W_1$ and $W_2$. Where $W_1$ and $W_2$ are the weighting matrices which are used for improving the estimation of $\Gamma_i X_f$ .We compute the SVD of $W_1.O_i.W_2$:

$$
\begin{aligned} W_1.O_i.W_2 &= [U_1 \quad U_2] \begin{bmatrix} S_1 & 0 \\ 0 & 0 \end{bmatrix} \begin{bmatrix} V_1^T \\ V_2^{\perp T} \end{bmatrix} \\ W_1.O_i.W_2 &= U_1.S_1.V_1^T \end{aligned} \tag{13}
$$

$S_1$ is a diagonal matrix formed by $n$ singular values different from zero. The order of the system is then $n$.

$$O_i = \Gamma_i X_f$$
$$\Gamma_i = W_1^{-1} U_1 S_1^{1/2} T$$
$$X_f W_2 = T^{-1} S_1^{1/2} V_1^T$$
$$X_f = \Gamma_i^{\dagger} O_i$$

$T \in R^{n \times n}$ is a non-singular similarity transformation matrix.

**Determination of Matrices**

The matrix $C$ is extracted directly from the first $l$ rows of $\Gamma_i$.The matrix $A$ is determined from the shift structure of $\Gamma_i$. Denoting

$$\underline{\Gamma_i} A = \overline{\Gamma}_i \ \rightarrow \ A = \underline{\Gamma_i}^{\dagger} \overline{\Gamma}_i \tag{14}$$

Where $\underline{\Gamma_i}$ is the matrix $\Gamma_i$ without the last $l$ rows:

$$\underline{\Gamma_i} = \begin{bmatrix} C \\ CA \\ CA^2 \\ \vdots \\ CA^{i-2} \end{bmatrix}$$

And $\overline{\Gamma}_i$ is $\Gamma_i$ without the first $l$ rows:

$$\overline{\Gamma}_i = \begin{bmatrix} CA \\ CA^2 \\ \vdots \\ \vdots \\ CA^{i-1} \end{bmatrix}$$

While multiplying (15) at left by $\Gamma_i^{\perp}$ and at right by $U_f^{\dagger}$ one obtains then:

$$Y_f = \Gamma_i X_f + H_i U_f \tag{15}$$

$$\Gamma_i^{\perp} Y_f U_f^{\dagger} = \Gamma_i^{\perp} \Gamma_i X_f U_f^{\dagger} + \Gamma_i^{\perp} H_i U_f U_f^{\dagger} \tag{16}$$

And knowing that the product $\Gamma_i^{\perp} \Gamma_i$ is null, it results in

$$\Gamma_i^{\perp} Y_f U_f^{\dagger} = \Gamma_i^{\perp} H_i \tag{17}$$

Denote

$$L = \Gamma_i^{\perp} \tag{18}$$

$$M = \Gamma_i^{\perp} Y_f U_f^{\dagger} \tag{19}$$

The equations (18) and (19) become

$$M = L H_i \tag{20}$$

A system of equations which are function of $B$ and $D$ is resolved by a linear regression algorithm [26, 27].

## 3. Plasma Open-loop Response Identification

### 3.1 Model Identification

For linear and nonlinear system identification, there are different methods. Before selection of the method, linearity or nonlinearity of system behavior must be determined. The coherence function provides a quantification of deviations from linearity in the system which lies between the input and output signals.

The coherence (sometimes called magnitude-squared coherence) between two signals x(t) and y(t) is a real-valued function that is defined as [31]:

$$G_{xy}(f) = \frac{\left|G_{xy}(f)\right|^2}{G_{xx}(f)G_{yy}(f)} \qquad (21)$$

Where $G_{xy}(f)$ is the cross-spectral density between x and y, and $G_{xx}(f)$ and $G_{yy}(f)$ the auto spectral density of x and y, respectively. The magnitude of the spectral density is denoted as $|G|$. The given restrictions noted above (ergodicity, linearity) the coherence function estimates the extent to which y(t) may be predicted from x(t) by an optimum linear least squares function.

This function has been calculated for the Damavand tokamak data and it shows nonlinearity of system. As an example, the magnitude squared coherence estimate of input(Icz) and output (plasma vertical position) signals for the shot (2015/27/05 Shot# 83) is shown in figure 5.

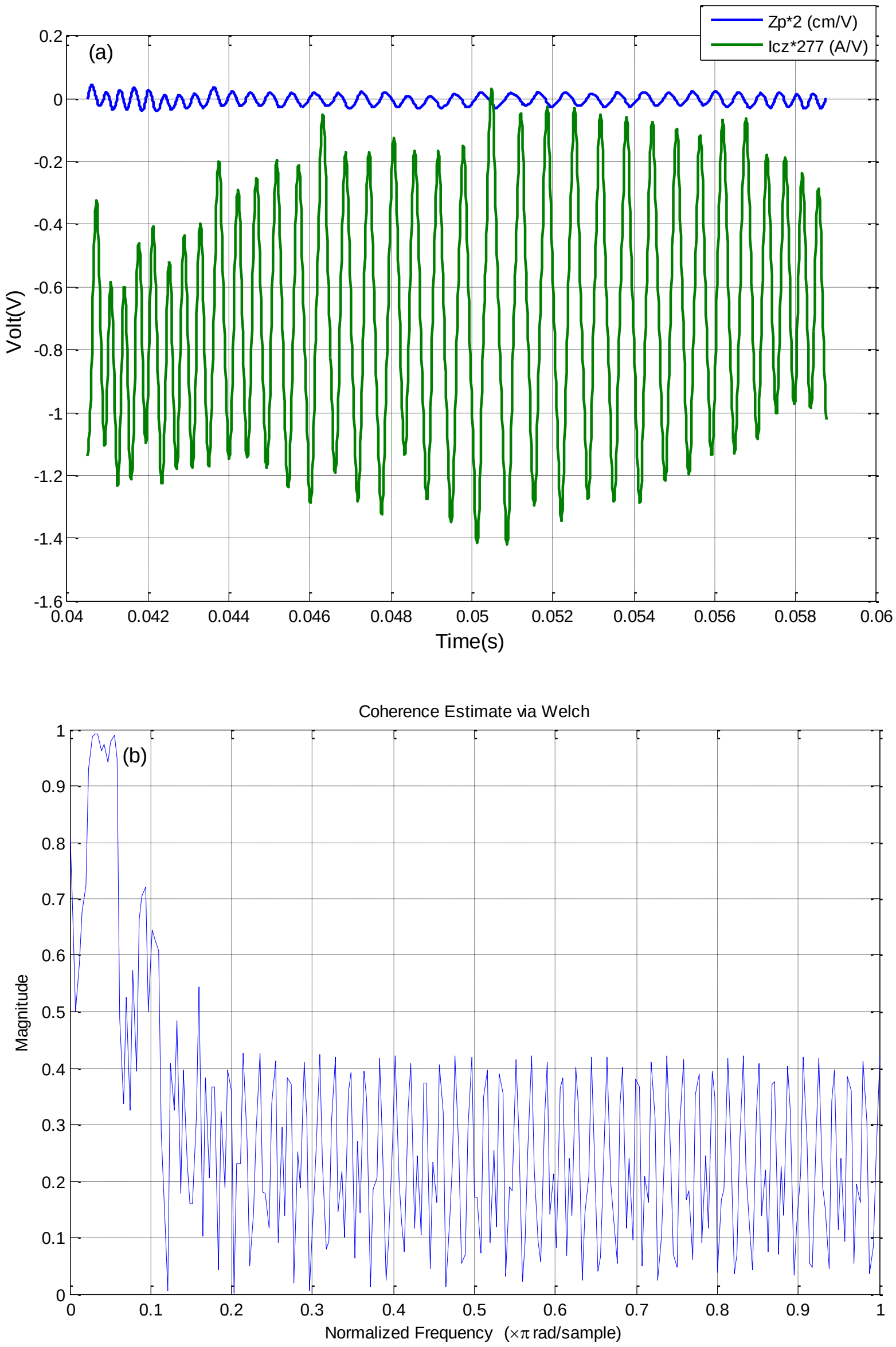

FIG. 5. (a) input(Icz) and output (plasma vertical position) signals for the shot (2015/27/05 Shot# 83). (b) the magnitude squared coherence estimate between input and output of the Damavand tokamak for the shot (2015/27/05 Shot# 83).

Nonlinear system identification (neural network) can be done similar to last works on the Damavand tokamk [10, 20]. The biggest criticism of the neural network models is that the produced models are completely opaque and usually cannot be written down or analyzed. It is therefore very difficult to either know what is causing that, or to analyze the model, or to compute dynamic characteristics from the model. But the aim of this work is a linear approximation of a nonlinear tokamka system that is valid in a small region around the operating point.

Linearization is useful in model analysis and control design applications. Exact linearization of the nonlinear Damavand tokamak model produces linear state-space, transfer-function, or zero-pole-gain equations that can be used to[32]:

- Plot the Bode response of the model.
- Evaluate loop stability margins by computing open-loop response.

- Analyze and compare plant response near different operating points.
- Design linear controller (Classical control system analysis and design methodologies require linear, time-invariant models.)
- Analyze closed-loop stability.
- Measure the size of resonances in frequency response by computing closed-loop linear model for control system.
- Generate controllers with reduced sensitivity to parameter variations and modeling errors (Robust Control).

Continuous-time nonlinear differential equations can be written in this form:

$$x^{\cdot}(t) = f(x(t), u(t), t) \;\; y(t) = g(x(t), u(t), t)$$

In these equations, $x(t)$ represents the system states, $u(t)$ represents the inputs to the system, and $y(t)$ represents the outputs of the system.

A linearized model of this system is valid in a small region around the operating point $t = t_0, x(t_0) = x_0, u(t_0) = u_0$, and $y(t_0) = g(x_0, u_0, t_0)$

To represent the linearized model, define new variables centered about the operating point: $\delta x(t) = x(t) - x_0, \delta u(t) = u(t) - u_0, \delta y(t) = y(t) - y_0$

The linearized model in terms of $\delta x$, $\delta u$, and $\delta y$ is valid when the values of these variables are small: $\delta x^{\cdot}(t) = A\delta x(t) + B\delta u(t), \delta y(t) = C\delta x(t) + D\delta u(t)$

Choosing the right operating point for linearization is critical for obtaining an accurate linear model. The linear model is an approximation of the nonlinear model that is valid only near the operating point at which the model is linearized [33].

In this section, a linear model is identified for plasma vertical position, based on the numerical algorithms for subspace state space system identification (N4SID) structure that explained in section 2.2. This model has been employed for estimation of the Damavand tokamak simulator parameters.

Figure 6 shows general view of the closed loop system in which $Z_p$ is vertical position and $\hat{Z}_p$ is model output. Inputs of the model are the controller current ($I_{cz}$) and the equilibrium current ($I_{eq}$). In the last identification study on the Damavand tokamak [10],$I_{eq}$ was not used as an input for identification process. But here for comparing to physical model, it has been introduced as an input.

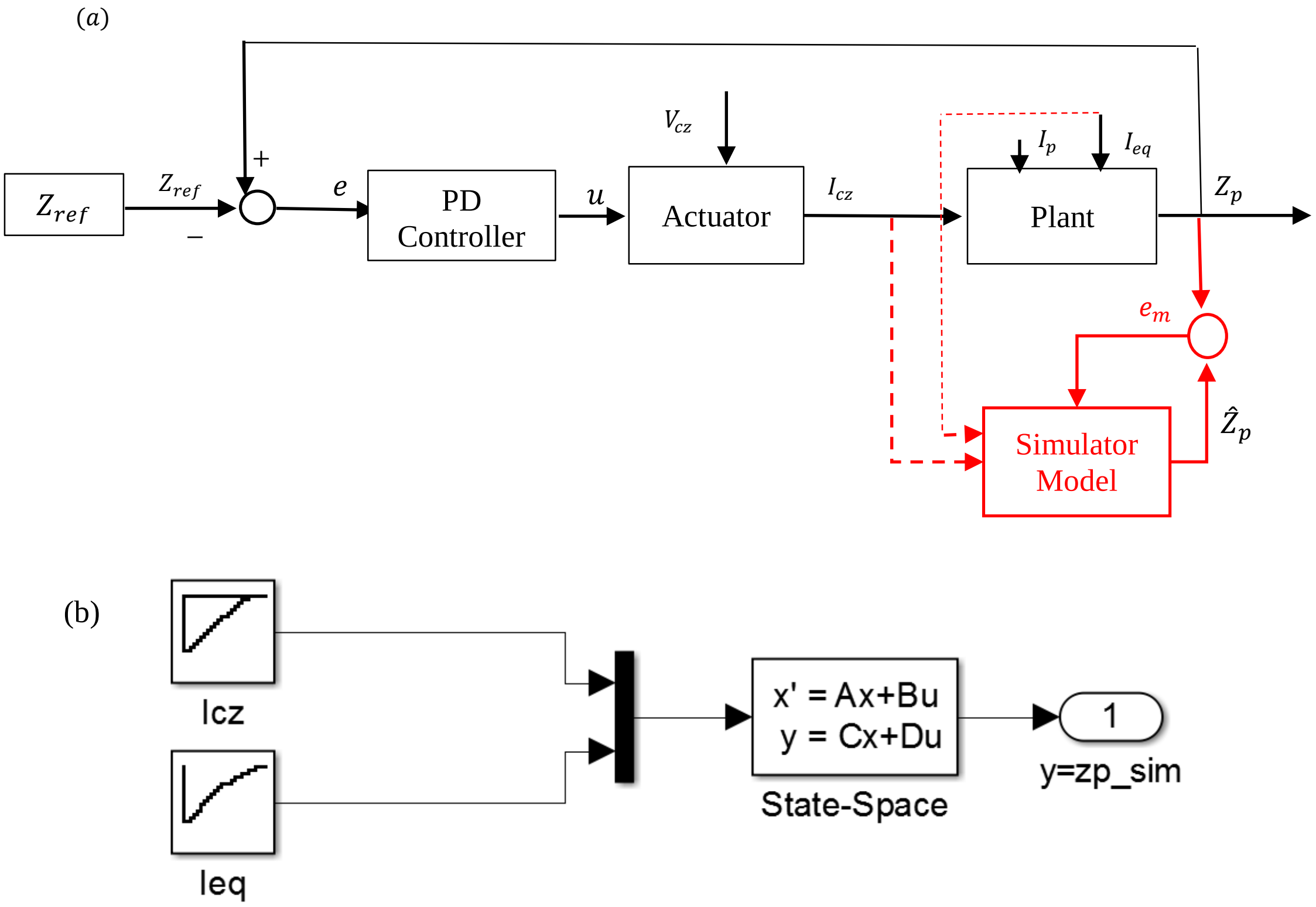

FIG. 6. (a) Block diagram of the closed loop system and the general structure of simulator model. (b) Simulator model structure used for identification.

The accurate value of delay is a vital parameter in system identification. Output delay of $I_{cz}$ and $I_{eq}$ was 200µs [10].

Data of different shots are gathered and some of them are selected for model testing. The sampling time is experimentally determined. Different sampling times are examined to find the best sampling time for the model. The simulated response of a dynamic system model, superimposed over validation data (Figure 5.a), has been plotted in Figure 7 for comparison. The plot also displays the normalized root mean square (NRMSE) measure of the goodness of fit. The observability matrix of the subspace model has been calculated and it is full rank that seems the identified model is observable. Also, the controllability matrix has been calculated and it is full rank that means the model is controllable. If identification range around operating point be reduced, better NRMSE can be obtained. As an example, the calculated matrices A, B and C (matrix D has been considered zero) for N4SID identification around operating point mentioned in Figure 7 are:

$$operating\ point \begin{cases} Ip = 3.533 * 10\ (\frac{kA}{V}) \\ Icz = -0.245 * 277\ (\frac{A}{V}) \\ Ieq = 1.323 * 500\ (\frac{A}{V}) \\ Zp = 0.003541 * 2\ (\frac{cm}{V}) \end{cases}, \qquad A = \begin{bmatrix} 0.8503 & -0.08158 \\ 0.09313 & 0.8754 \end{bmatrix},$$

$$B = \begin{bmatrix} 0.001911 & 0.04813 \\ 0.003548 & 0.03437 \end{bmatrix}, C = [\,0.02497 \quad -0.008559], D = 0.$$

(a)

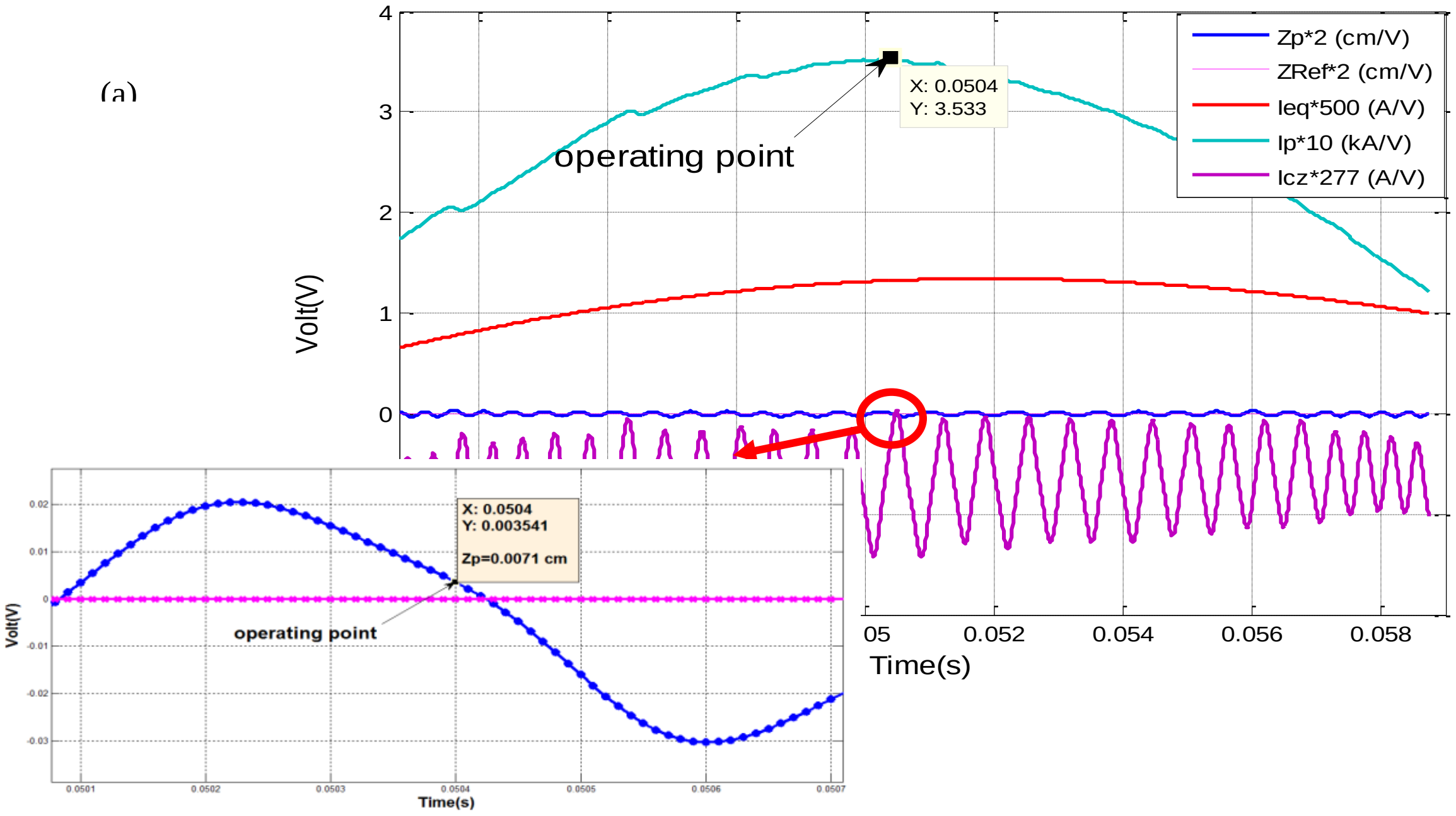

(b)

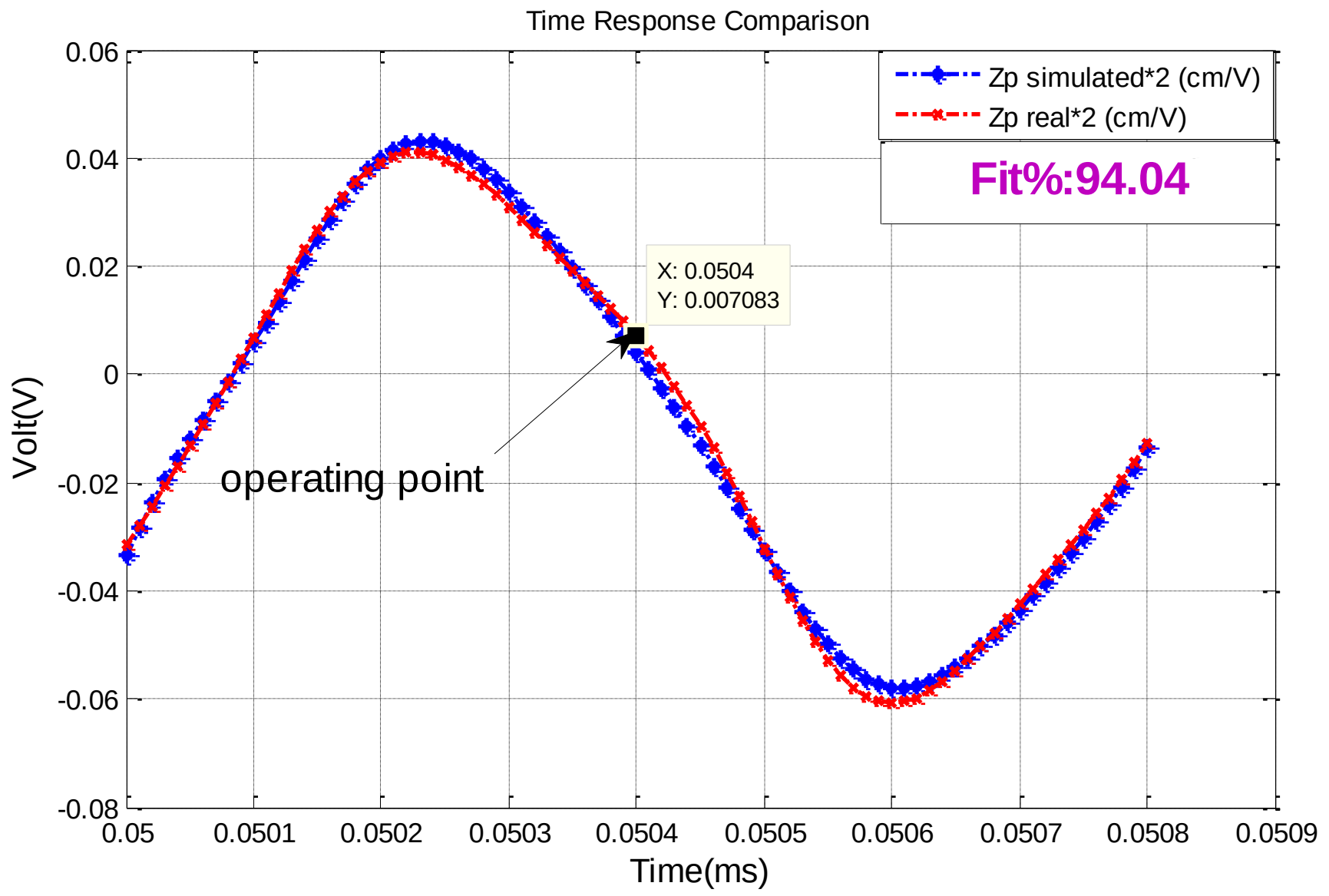

FIG. 7. (a) Experiment data for validation of subspace model around operating point (shot (2015/27/05 Shot# 83)). (b) Simulated response of a dynamic Damavand tokamak model (around operating point), superimposed over validation data (Figure 5.a (shot (2015/27/05 Shot# 83))).

## 3.2 Validation of Identified N4SID Model with Experimental Data and Physical Model

In this section, linearized tokamak plasma equilibrium model (a filamentary plasma model) based on calculated equilibrium parameter and decay index has been developed for comparison to identified model. This physical model is based on magnetic probe measurements [20]. In Figure 7, the results of identified model are illustrated together with the values of experimental output (2015/27/05 Shot# 83). The results show that maximum error in the plasma vertical position and the RMSE% are 0.0079 cm and 5.96%, respectively.

The result of physical modeling with condition of the shot (2015/27/05 Shot# 83) is shown in Figure 8 for comparison with test output of N4SID identification (Figure 7).The results show that maximum error in the plasma vertical position at flat top plasma current and the RMSE% for physical model are 0.2095 cm and 20.3515%, respectively. In Figure 9, comparison of output of N4SID identified simulator model with physical model output for shot (2015/27/05 Shot# 83) is shown. The result show that the identified model has very better goodness of fit than the physical model.

For validation of the identified model with experimental data, different shots with similar condition of shot (2015/27/05 Shot# 83) have been selected. Then the simulated outputs of plant around operating point (mentioned in Figure 7) for each shot have been calculated. As examples, simulation results of shot (2015/27/05 Shot# 81) and shot (2015/27/05 Shot# 85) are shown in Figures 10 and 11 respectively. For shot (2015/27/05 Shot# 81) maximum error in the plasma vertical position at flat top plasma current and the RMSE% for identified output are 0.0032cm and 2.0709%, respectively and for (2015/27/05 Shot# 85) are 0.0035 cm and 3.6798%.

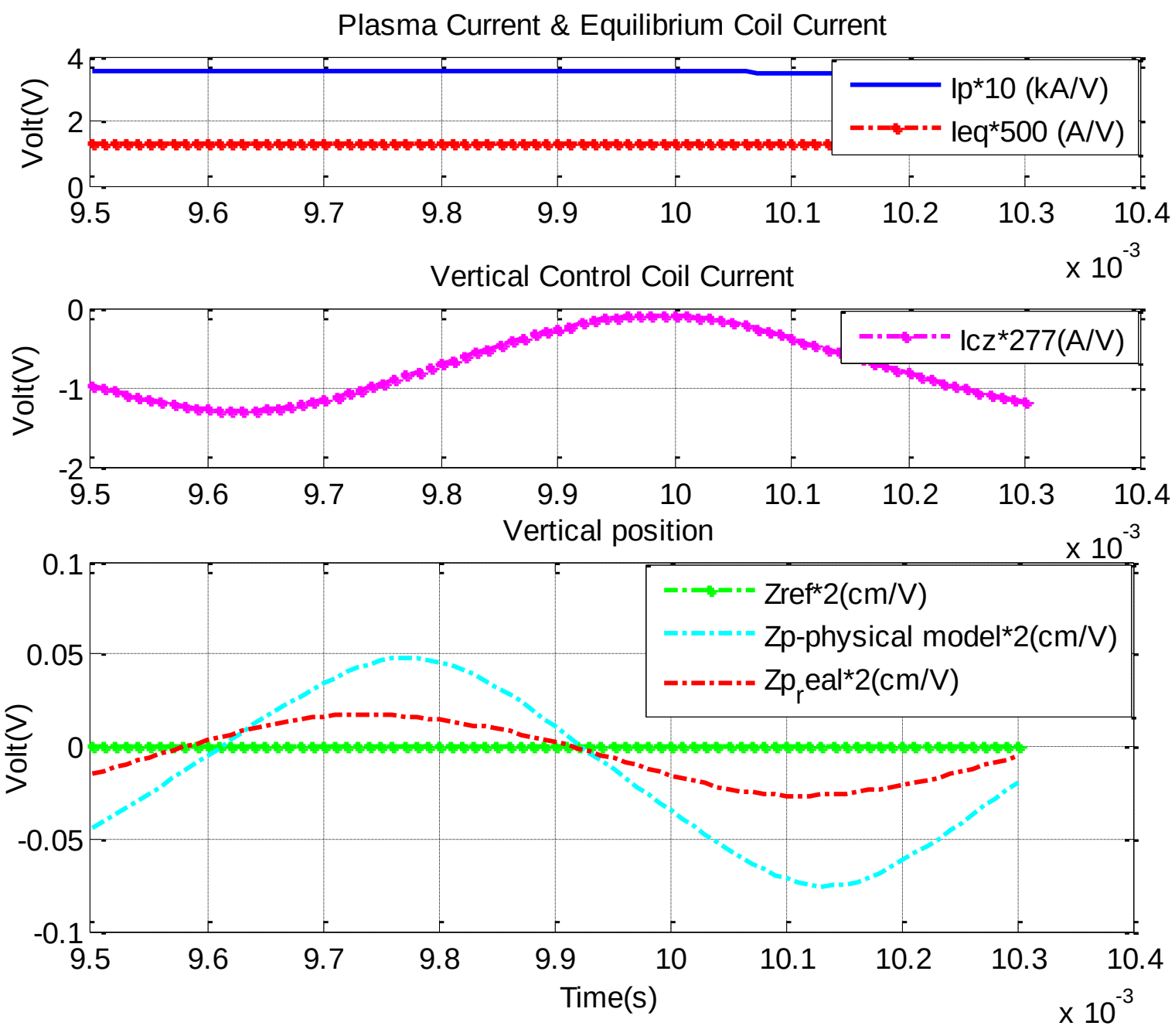

FIG. 8. Comparison of output of physical model with experimental output for test (2015/27/05 Shot# 83). (a) Input of system (plasma current and equilibrium coil current). (b) Input of system (vertical coil current). (c) Output of real system and physical model of system.

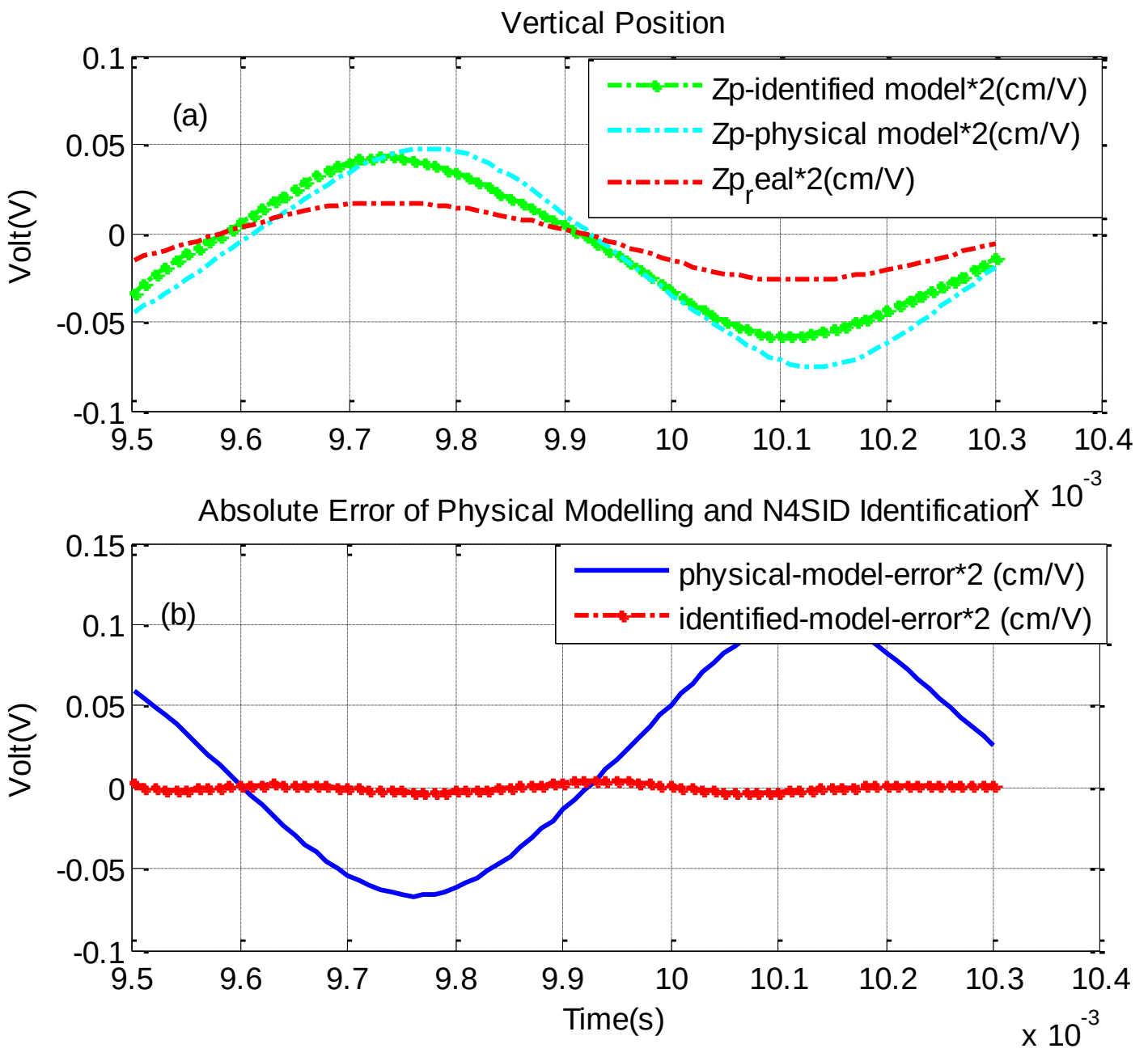

FIG. 9. Comparison of output of N4SID identified simulator model with physical model output for test (2015/27/05 Shot# 83). (a) Output of real system and identified model of system and physical model output(b) Absolute error of identified model test and physical modelling

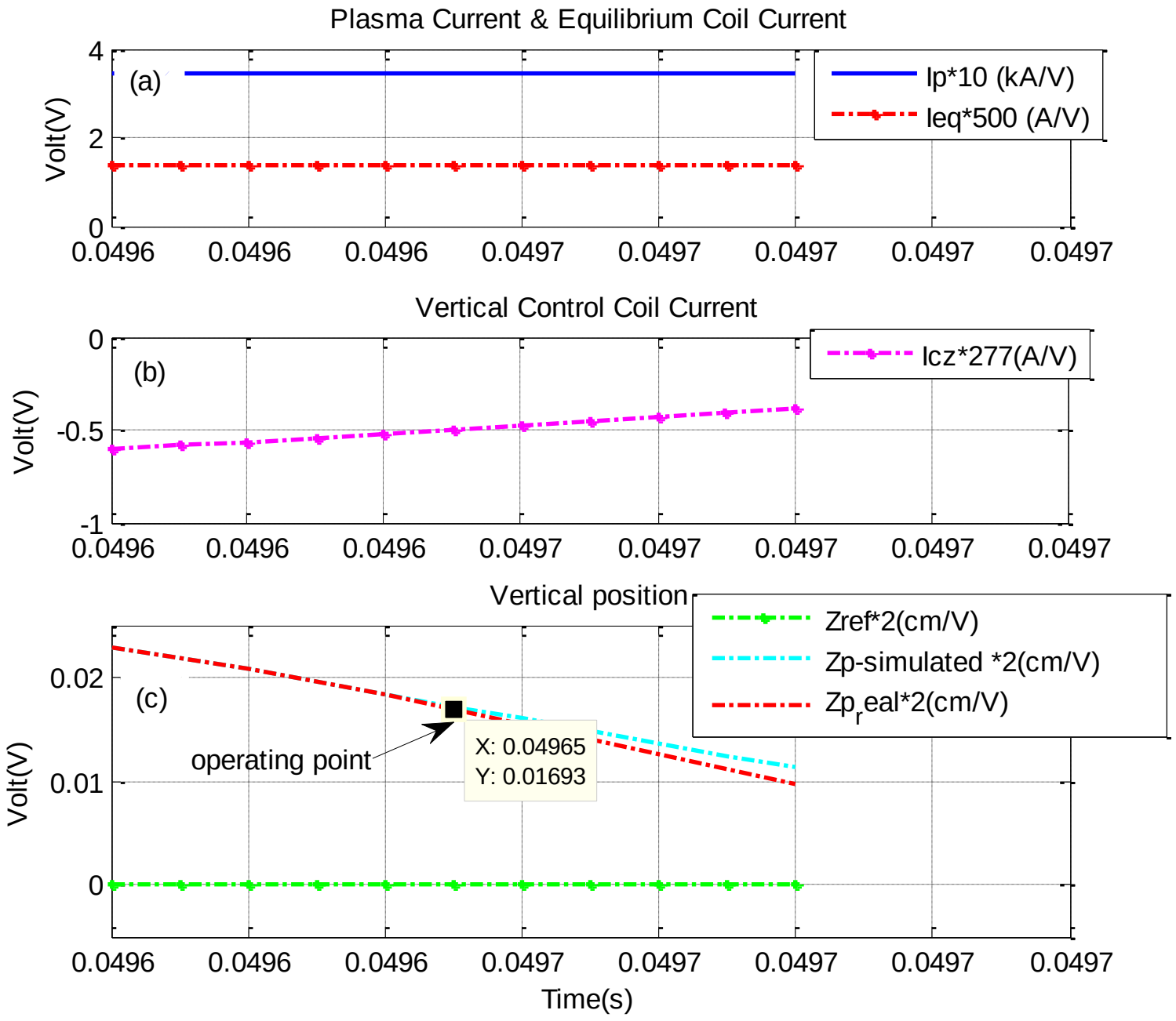

FIG. 10. Simulated response of shot (2015/27/05 Shot# 81) inputs with N4SID identified model

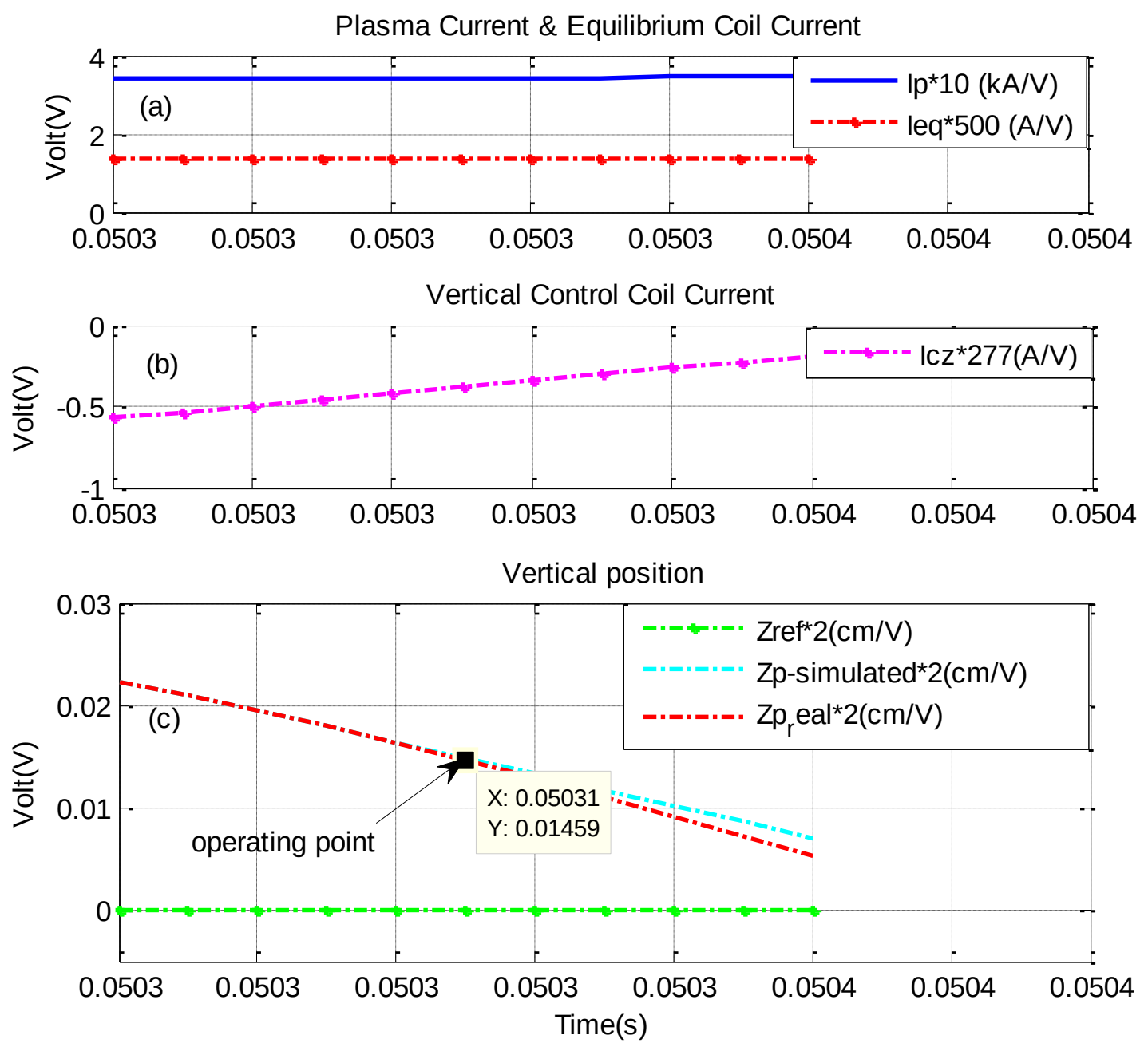

FIG. 11. Simulated response of shot (2015/27/05 Shot# 85) inputs with N4SID identified model.

## 4. Conclusions

The objective of this paper was to create linear model which is suitable for controller design and which may be used to validate different models derived from physical principles. In this paper we described N4SID subspace system identification technique to create a linear model of the response of a Damavand tokamak plasma to voltages applied to the poloidal field coils. The obtained model is MISO. The ability of identified model to predict system behaviour was tested by different shots. Some examples are shown in Figures 7, 10 and 11.The result show that RMSE% of simulated shots with the identified model by N4SID algorithm, are less than 10%. For an example the maximum error in the plasma vertical position simulation and the %RMSE of modeling are 0.0032 cm and 2.0709%, respectively for particular shot (2015/27/05 Shot# 81).

Also, the performance of identified model is compared with physical model. This physical model is based on magnetic measurement and the algebraic equation defining the vertical position in this model, is based on instantaneous force-balance[20]. The results show that the identified model has very better goodness of fit than physical model and can predict system behaviour more accurately.

## References

1. Artsimovich, L.A., *Tokamak devices.* Nuclear Fusion, 1972. **12**(2): p. 215.
2. Liu, L., et al., *Controllability study of EAST plasma vertical instability and improvement in future.* Fusion Engineering and Design, 2014(0).
3. Favez, J.Y., et al., *Improving tokamak vertical position control in the presence of power supply voltage saturation.* Plasma Physics and Controlled Fusion, 2005. **47**(10): p. 1709.
4. Ljung, L., *System identification : theory for the user*. 1987, Englewood Cliffs, NJ: Prentice-Hall.

5. Moreau, D., et al., *A two-time-scale dynamic-model approach for magnetic and kinetic profile control in advanced tokamak scenarios on JET.* Nuclear Fusion, 2008. **48**(10): p. 106001.
6. Coutlis, A., et al., *Frequency response identification of the dynamics of a Tokamak plasma.* Control Systems Technology, IEEE Transactions on, 2000. **8**(4): p. 646-659.
7. Yoshida, M., et al., *Momentum transport and plasma rotation profile in toroidal direction in JT-60U L-mode plasmas.* Nuclear Fusion, 2007. **47**(8): p. 856.
8. Moreau, D., et al., *Plasma models for real-time control of advanced tokamak scenarios.* Nuclear Fusion, 2011. **51**(6): p. 063009.
9. Wehner, W., et al. *Data-driven modeling and feedback tracking control of the toroidal rotation profile for advanced tokamak scenarios in DIII-D.* in *Control Applications (CCA), 2011 IEEE International Conference on*. 2011.
10. Rasouli, H., C. Rasouli, and A. Koohi, *Identification and control of plasma vertical position using neural network in Damavand tokamak.* Q2 Review of Scientific Instruments, 2013. **84**(2): p. 023504-023504-12.
11. Rasouli, H. and A. Fatehi, *Design of set-point weighting PIλ+ Dμ controller for vertical magnetic flux controller in Damavand tokamak.* Review of Scientific Instruments, 2014. **85**(12): p. 123508.
12. Rasouli, H., A. Fatehi, and H. Zamanian, *Design and implementation of fractional order pole placement controller to control the magnetic flux in Damavand tokamak.* Review of Scientific Instruments, 2015. **86**(3): p. 033503.
13. Bin, W., et al., *First operations with the new Collective Thomson Scattering diagnostic on the Frascati Tokamak Upgrade device.* Journal of Instrumentation, 2015. **10**(10): p. P10007.
14. Czarski, T., et al., *Serial data acquisition for the X-ray plasma diagnostics with selected GEM detector structures.* Journal of Instrumentation, 2015. **10**(10): p. P10013.
15. Rasouli, C., et al., *Runaway electron energy measurement using hard x-ray spectroscopy in "Damavand" tokamak.* Review of Scientific Instruments, 2009. **80**(1): p. 013503.
16. Rasouli, C., et al., *Study of runaway electrons using dosimetry of hard x-ray radiations in Damavand tokamak.* Review of Scientific Instruments, 2014. **85**(5): p. 053509.
17. Khayrutdinov, R.R. and V.E. Lukash, *Studies of Plasma Equilibrium and Transport in a Tokamak Fusion Device with the Inverse-Variable Technique.* Journal of Computational Physics, 1993. **109**(2): p. 193-201.
18. *System identification (2nd ed.): theory for the user*, ed. L. Lennart. 1999: Prentice Hall PTR. 609.
19. *TOKAMAK "DAMAVAND" Technical Document*. 1994, AEOI: MOSCOW. p. 67.
20. Darestani Farahani, N. and F. Abbasi Davani, *Experimental Determination of Some Equilibrium Parameter of Damavand tokamak by Magnetic Probe Measurements for Representing a physical Model for Plasma Vertical Movement.* Review of Scientific Instruments, 2015. **86**(10).
21. Yu, J.-L., *A novel subspace tracking using correlation-based projection approximation.* Signal Processing, 2000. **80**(12): p. 2517-2525.
22. Niederlinski, A., *THEORY AND PRACTICE OF RECURSIVE IDENTIFICATION, Lennart Ljung and Torsten Soderstrom, The MIT Press, Cambridge, Massachusetts, London, England, 1983. Price: £45.00. No. of pages: 529.* Optimal Control Applications and Methods, 1985. **6**(1): p. 71-72.
23. Xu, B.-C. and X.-L. Liu, *Identification algorithm based on the approximate least absolute deviation criteria.* International Journal of Automation and Computing, 2012. **9**(5): p. 501-505.
24. Guo, Y. and J. Tan, *A time-domain model-based method for the identification of multi-frequency signal parameters.* Journal of Instrumentation, 2014. **9**(06): p. P06019.

25. Gustafsson, T. and M. Viberg. *Instrumental variable subspace tracking with applications to sensor array processing and frequency estimation*. in *Statistical Signal and Array Processing, 1996. Proceedings., 8th IEEE Signal Processing Workshop on (Cat. No.96TB10004*. 1996.
26. Van Overschee, P. and B. De Moor, *N4SID: Subspace algorithms for the identification of combined deterministic-stochastic systems.* Automatica, 1994. **30**(1): p. 75-93.
27. Overschee, P.v. and B.L.R.d. Moor, *Subspace identification for linear systems: theory, implementation, applications*. 1996, Boston: Kluwer Academic Publishers. xiv, 254 p.
28. Liao, Z., et al., *Subspace identification for fractional order Hammerstein systems based on instrumental variables.* International Journal of Control, Automation and Systems, 2012. **10**(5): p. 947-953.
29. Borjas, S.D.M. and C. Garcia, *Subspace identification for industrial processes.* TEMA (São Carlos), 2011. **12**: p. 183-194.
30. Hachicha, S., M. Kharrat, and A. Chaari, *N4SID and MOESP Algorithms to Highlight the Ill-conditioning into Subspace Identification.* International Journal of Automation and Computing, 2014. **11**(1): p. 30-38.
31. Piersol, A.G. and J.S. Bendat. *Random data analysis and measurement procedures*. 2013; Available from: http://rbdigital.oneclickdigital.com.
32. *Linearizing Nonlinear Models*. 1994-2015, The MathWorks, Inc.
33. Verdult, V., *Nonlinear System Identification: A State-Space Approach*, in *Faculty of Applied Physics*. 2002, University of Twente: Twente University Press.